\documentclass[useAMS,usenatbib,usegraphicx,twocolumn]{mn2e}
\usepackage[figuresright]{rotating}

\title[Orbital structure in N-Body models of barred galaxies]{Orbital structure in barred galaxies}
\author[N. Voglis, M. Harsoula, and G. Contopoulos]
       {N. ~Voglis, \thanks{This work started by Dr. N. Voglis with Dr. M. Harsoula. It was very
sad that Dr. Voglis passed away suddenly on 9/2/2007. After that Dr.
G. Contopoulos continued this work and completed this paper in
collaboration with Dr. M. Harsoula.} M. ~Harsoula and G. ~Contopoulos\\
        Research Center for Astronomy and Applied Mathematics,
           Academy of Athens, Soranou Efesiou 4, GR-115 27 Athens, Greece\\
           e-mail: mharsoul@academyofathens.gr,
           gcontop@academyofathens.gr}
\date{Released 2007 April 12}
\pagerange{\pageref{firstpage}--\pageref{lastpage}} \pubyear{2007}

\def\LaTeX{L\kern-.36em\raise.3ex\hbox{a}\kern-.15em
    T\kern-.1667em\lower.7ex\hbox{E}\kern-.125emX}

\begin{document}

\maketitle

\label{firstpage}

\begin{abstract}
We study the orbital structure of a self-consistent N-body
equilibrium configuration of a barred galaxy constructed from
cosmological initial conditions. The value of its spin parameter
$\lambda$ is near the observed value of our Galaxy $\lambda=0.22$.
We classify the orbits in regular and chaotic using a combination of
two different methods and find 60\% of them to be chaotic. We
examine the phase space using projections of the 4D surfaces of
section for test particles as well as for real N-body particles. The
real particles are not uniformly distributed in the whole phase
space but they avoid orbits that do not support the bar. We use
frequency analysis for the regular orbits as well as for the chaotic
ones to classify certain types of orbits of our self-consistent
system. We find the main resonant orbits and their statistical
weight in supporting the shape of the bar and we emphasize the role
of weakly chaotic orbits in supporting the boxiness at the end of
the bar.

\end{abstract}

\begin{keywords}
galaxies:structure,kinematics and dynamics.
\end{keywords}

\section{Introduction}

The type of orbits that appear in various types of galaxies have
been studied extensively up to now. (For a review see Contopoulos
2002). However relatively few studies have been made on the orbital
structure of N-body systems (see for example \citet{b23},
\citet{b19}, \citet{b20},
 \citet{b15}, \citet{b5}, \citet{b18}, \citet{b6}). In
this paper we consider a N-body model of a triaxial barred galaxy.
Its spin parameter is relatively large, comparable to the spin
parameter of our Galaxy. Our study aims to find the main types of
orbits in such a galaxy. We emphasize the chaotic orbits that are
60\% of the totality, in contrast with non-rotating N-body systems
where the chaotic orbits are only 30\%. The chaotic orbits occupy
most of the space outside corotation. However the chaotic orbits are
not completely mixed. They are separated for long times by cantori.
Relatively few chaotic orbits escape from the system during a Hubble
time.

There are also many ordered orbits of various resonant types, mainly
inside the bar. The topology of the orbits is governed by the
resonances between the rotational and the epicyclic frequencies. The
third dimension plays a minor role on the type of orbits.

A comparison between the N-body orbits and the orbits of test
particles in the same potential shows that real orbits avoid
systematically certain types of resonant orbits, occupying
preferentially orbits that support the bar. In fact models
consisting of particles uniformly distributed initially are not
self-consistent.

The paper is organized as follows: In section 2 we describe the
method used for constructing a realistic bar model and in section 3
we describe the characteristic features of the bar. Section 4
discusses the level of chaos. Section 5.1 describes the various
types of orbits by using surfaces of section and section 5.2
introduces a frequency analysis of the orbits. Finally in section 6
we summarize our conclusions.

\section[]{Cosmological Initial Conditions for creating a bar-like galaxy}

For creating the final configuration of a bar-like galaxy we have
used two different N-body codes: a tree code developed by
\citet{b12} and a conservative technique code designed by
\citet{b1}. Initially 5616 particles of equal mass are positioned in
a cubic grid limited by a sphere and having a total mass equal to
the mass of a galaxy.The system expands initially according to an
Einstein-de Sitter model universe and then we impose a density
perturbation with a spherical profile and a power law dependence on
the radius $r$, i.e.

\begin{equation}
s(r)=\frac{\delta\rho}{\rho}\approx \frac{1}{r^{(n+3)/2}}
\end{equation}

 Such a profile is consistent with a power-law spectrum
of the density perturbation in the early Universe. Here we set
 $n=-2$, a value that is in the range predicted by the Cold Dark
 Matter scenario for galactic mass scales (Davis et al. 1985).
 Further details for these "quiet initial conditions" can be found in \citet{b27}.
The unit length is 1$kpc$, and the unit of time $t_{unit}$=1$Myr$
while the total mass is taken as $10^{12} M_{\odot}$. Using the tree
code, we find that the violent relaxation process has been completed
after $t=4000 tunits$. Then we split each particle into 27 particles
arranged in a small cubic grid, while the sum of these children
particles conserves the mass, energy and angular momentum of the
parent particle. The new configuration consists of about 1.4x$10^5$
particles and is now put as initial condition for the new run using
the conservative technique code (hereafter c-t) for another 150
half-mass crossing times. The c-t code gives a smooth potential
containing 120 terms (monopole, quadrupole and triaxial). All the
details are given in \citet{b13}. In the same paper the time
evolution of the potential coefficients is shown in figs. 5a,c. They
seem to get stabilized fast and stay remarkably constant even for
times compared to the Hubble time. The final output gives a triaxial
elliptical galaxy. Its ellipticity is about E7 on the Y-Z plane and
about E5 on the X-Y plane, in a radius that corresponds to 2 half
mass radii (hereafter $hmr$).

 Since our aim is to create a system that is rotating
around its smallest axis (i.e. the z-axis)we simply rotate the
projections of the velocity vectors of all the particles on the X-Y
plane to become perpendicular to their position vectors with a
counterclockwise sense of rotation, and therefore we give the
maximum possible rotation on this plane. This mechanism has already
been used in \citet{b31}. Then we let the system evolve using the
c-t code again, for another 250 dynamical times and calculate the
"spin parameter" $\lambda$ (Peebles ,1969) as a function of the
radius. The spin parameter is calculated by the following formula:

\begin{equation}
\lambda=\frac{J|E|^{1/2}}{GM^{5/2}}
\end{equation}
where G is the gravitational constant. The angular momentum $J$, the
bounding energy $E$ (in the inertial frame), and the mass $M$ are
measured as cumulative quantities along cylinders on the X-Y plane,
having the Z-axis as their main axis.

\begin{figure}
\includegraphics*[width=8.cm]
{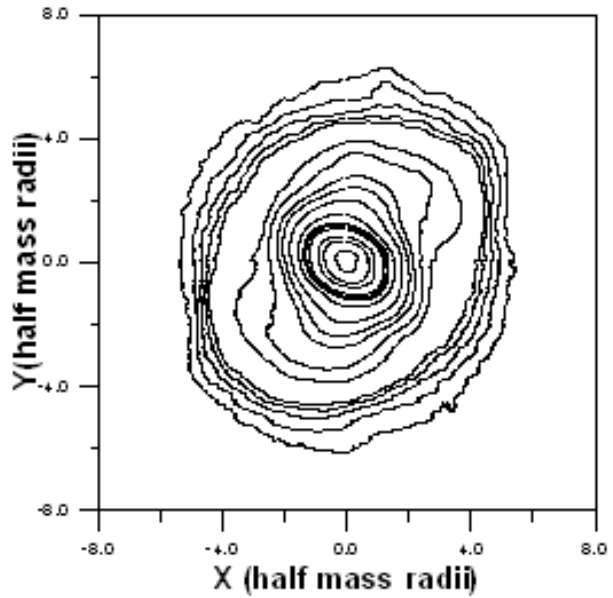}
 \caption{A snapshot of the isodensity contours at time=12$hmct$}
\end{figure}

At a time $t=250 tunits$ the value of $\lambda$ has increased
dramatically in the inner parts of the barred galaxy and stays close
to $\lambda=0.22$ over most of the extent of the system. This value
corresponds to the maximum value of $\lambda$ for a rotationally
supported galaxy and is close to the one of our Galaxy, as it was
calculated by \citet{b34}.

In the c-t code we use a new $runit$ that corresponds to the
limiting radius of the whole bound system i.e. one $runit$ is
approximately 120 kpc in real units and a new $tunit$ that
corresponds to the half mass crossing time of the system (hereafter
$hmct$). Since a Hubble time is $\approx300hmct$ (see \citet{b13}
for details), then $1hmct \approx45,6 Myrs$.

\section{Characteristics of the bar}

\begin{figure}
\includegraphics*[width=7.0cm,angle=-90]
{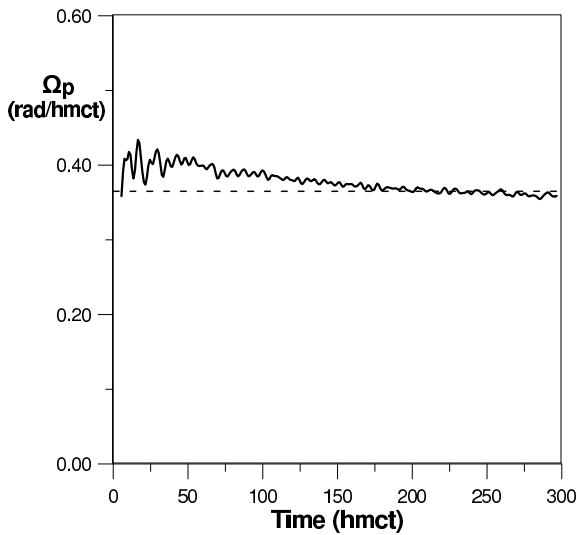}
 \caption{The time evolution of
 $\Omega_ p$ of the bar, in radians per half mass crossing time(hmct)}
\end{figure}

In order to calculate the angular velocity of the pattern (hereafter
$\Omega_p$) of the system, we must find the time evolution of the
angle between the great axis of inertia of the rotating system and a
fixed axis, that is the Y-axis of the inertial frame of reference.
However, by observing the time evolution of the isodensities on the
rotating plane, we notice the existence of a differential rotation
of the pattern, i.e. the inner parts seem to rotate faster that the
outer parts at the beginning of the c-t run. This lasts up to a
certain time, until a single $\Omega_p$ is established throughout
the radius of the whole configuration. In Fig. 1 a snaphot of the
isodensities on the X-Y plane is shown at 12 half mass crossing
times ($hmct$) from the beginning of the c-t run. We notice a
trailing spiral structure, as the galaxy is rotating as a whole
counterclockwise. This spiral structure seems to travel outwards
before it vanishes later on. Therefore angular momentum is
transferred to the material outwards. This is a well known mechanism
that makes the bar grow stronger and slow down (Lynden-Bell and
Kalnajs, 1972). These authors have shown that density waves in the
form of temporary spiral structures carry angular momentum from the
inner parts to the outer parts. This transference is due to the
interaction of the bar perturbation of the gravitational field with
stars moving in resonant orbits particularly near the ILR and
corotation. They also proved that only in trailing spiral structures
do the gravity
 torques carry angular momentum outwards. Later on
\citet{b24} and \citet{b25} extended this analysis showing that
angular momentum transference can be considered as a particular type
of dynamical friction. N-body simulations have confirmed this
evolution scenario of bar-like galaxies (e.g. Athanassoula, 2003).

 In the present experiment this spiral structure vanishes after a short transient
period of our calculations i.e. at $t\approx 20 hmct$. However, in
experiments with a larger value of $\Omega_p$ these spiral
structures can survive even for times comparable to a Hubble time
(Voglis et al., 2006).

In Fig. 2 we plot the time evolution of the angular velocity in
units of $rad/hmct$ of the material inside 2.0 $rhm$. The
fluctuation of $\Omega_p$ at the beginning of the c-t run is related
to the differential rotation of the pattern which transfers angular
momentum outwards. This phenomenon creates spiral structures
transiently (see Fig. 1) and causes the formation of tidal torques
between the differentially rotating parts. Therefore the material
inside the bar can temporarily accelerate or decelerate, before the
whole system acquires the same $\Omega_p$. The very small
fluctuations after this transient time are less than $1^o/hmct$.
 We notice that the material of the bar seems to
slow down and stabilize its angular velocity, after
$t\approx150hmct$ to a value of $\Omega_p\approx21^o/hmct$. This
value of $\Omega_p$ is equivalent to
($2\pi$)/($17hmct$)=0.37$rad/hmct$ or to 8.25$km/(kpc.sec)$ (dashed
line in Fig. 2).

In order to study the orbital structure of this bar-like galaxy and
the level of chaos we need good statistics and therefore a second
multiplication of the particles is necessary. By considering only
the bound part of our galaxy (inside the radius 1.0 $runit$) we
split again each particle into 9 particles arranged in a small cubic
grid around the parent particle, while the sum of these children
particles conserves again the mass, energy and angular momentum of
the parent particle. The new configuration consists now of about
1.1x$10^6$ particles. We then let the system relax for another 50
half mass crossing times. After a short time period the system
reaches again an equilibrium and rotates with the same $\Omega_p$
pattern: $\Omega_p=0.37rad/hmct$ as a whole.

For studying the orbits that support the bar and determine the level
of chaos we take a certain snapshot, where the bar is parallel along
one of the main axes of the inertial frame of reference on the plane
of rotation, freeze the potential of the system and calculate the
orbits in the rotating frame of reference.

At 50 $tunits$ of the c-t run, after the new multiplication of the
particles, the bar of the galaxy has been aligned with the Y-axis of
the inertial system of reference. This is obvious in Fig. 3 where we
plot the isodensity contours on the X-Y plane.

In Fig. 4 the surface density profile is plotted along the main axis
of the bar (black line, Y-axis) and across the main axis of the bar
(grey line, X-axis). It is obvious that an exponential-like profile
can be fitted (thick line) in the first case almost until the limit
of the bar at $\approx 2hmr$, which is a well known property, in the
literature, for barred galaxies. The dashed line is the exponential
fit of the surface density across the main axis of the bar.

In Fig. 5 we plot the ellipticity=$1-b/a$, where $a$ is the
semi-major axis and $b$ is the semi-minor axis as a function of the
distance along the Y-axis. At this snapshot, the rotating frame of
reference coincides with the inertial frame of reference. The
calculation has been done using the isodensity contours of Fig. 3.
The ellipticity is close to E5 in the inner parts and drops to E1
near the end of the bar, while the isodensities become almost
spherical outside corotation.

\begin{figure}
\includegraphics*[width=7.5cm]
{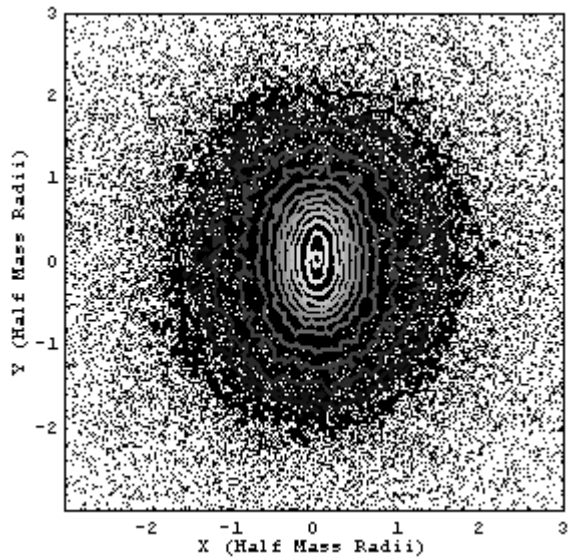}
 \caption{Isodensity contours on the X-Y plane at the end of the N-body run,
  together with the projections of the real particles
 of the galaxy.}
\end{figure}

\begin{figure}
\includegraphics*[width=7.5cm,angle=-90]
{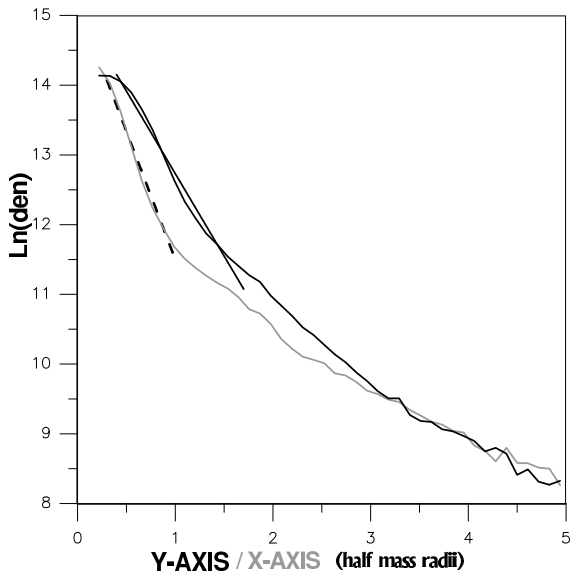}
 \caption{The surface density profile at a time=250thmass along the Y-axis (black line)
 and along the X-axis (grey line).}
\end{figure}

\begin{figure}
\includegraphics*[width=7.0cm,angle=-90]
{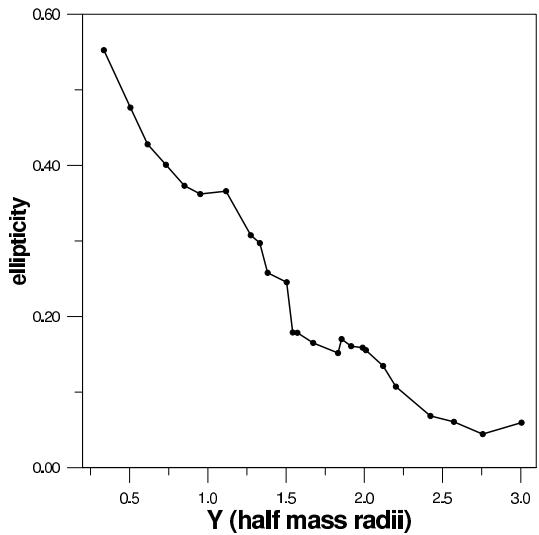}
 \caption{The ellipticity on the X-Y plane as a function of the semi-major axis of the bar.}
\end{figure}

The boxiness of the isophotes is a well known feature in barred
galaxies. This effect can be measured by the shape parameter $c$,
which can be determined from the equation of a generalized ellipse
(Athanassoula et al. 1990):

\begin {equation}
(|y|/a)^c+(|z|/b)^c=1
\end{equation}
where $a$ and $b$ are the semi-major and semi-minor axes (calculated
from the equidensity contours) and $c$ is the parameter describing
the shape of the generalized ellipse. For $c$=2 we obtain a standard
ellipse, for $c>2$ the shape approaches a parallelogram and for
$c<2$ we have a lozenge. In Fig. 6 we plot the shape parameter c as
a function of the semi-major axis of each isodensity contour of Fig.
3. From this figure we conclude that in our model, the inner and the
outer parts of the bar can be well fitted by ellipses while in a
region in between we have somewhat orthogonal-like isophotes. This
result agrees with the N-body experiments of \citet{b3}.

\begin{figure}
\includegraphics*[width=7.5cm,angle=-90]
{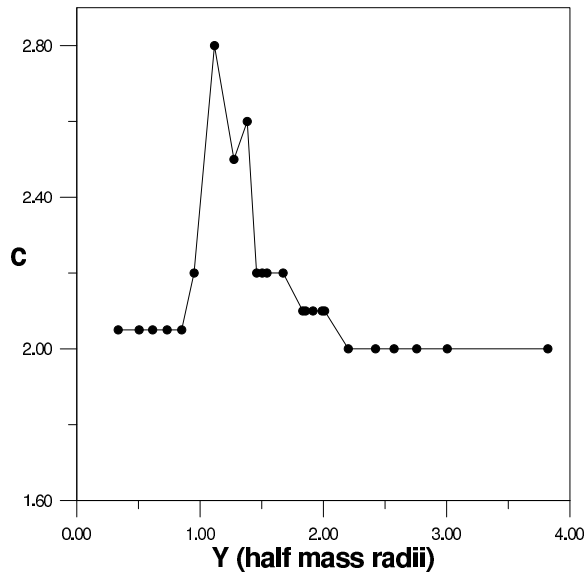}
 \caption{The shape parameter c, as a function of Y (semimajor axis).}
\end{figure}

\begin{figure}
\includegraphics*[width=7.0cm,angle=-90]
{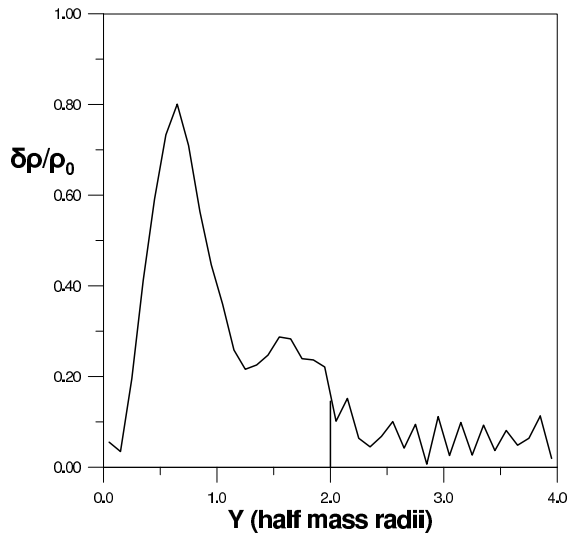}
 \caption{The bar perturbation $\delta \varrho/\varrho_0$
 as a function of y along the bar on the plane of rotation. The end of the bar is at $\approx2hmr$.}
\end{figure}

\begin{figure}
\includegraphics*[width=7.0cm,angle=-90]
{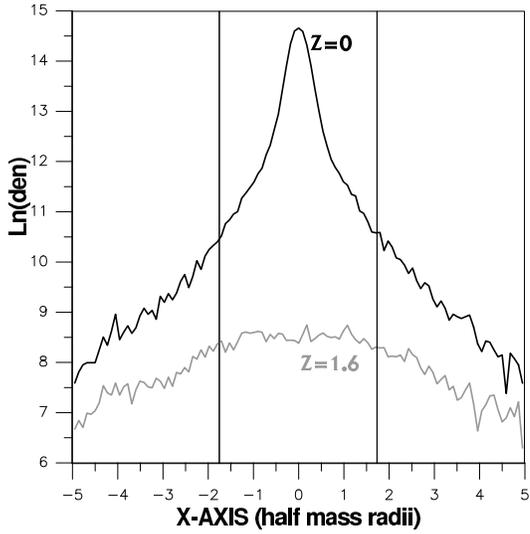}
 \caption{Surface density along the bar minor axis for z=0 (black line) and z=1.6$hmr$ (grey line).}
\end{figure}

An estimate of the perturbation of the bar is the value of $\delta
\varrho/\varrho_0$, where $\varrho_0$ is the mean density and
$\delta \varrho$ is the deviation from it along an annulus in a
certain distance from the center. The calculations are made by
projecting the particles on the plane of rotation (x-y plane) and
dividing the surface into annuli of equal width.  Then we also split
the xy plane in azimuth (with $d\theta=2*\pi/40$) into 40 successive
angles creating bins. Then we count
 the particles in each bin to calculate the density $\rho$. We plot the maximum amplitude
$\delta \varrho / \varrho_0 = (\varrho-\varrho_0) / \varrho_0$  in
every annulus as a function of y.
 By changing $d\theta$ from $d\theta=2*\pi/20$ up to $d\theta=2*\pi/50$ we have found optimal values, such that a
 further decrease of $d\theta$ does not change the maximum amplitudes.
 In Fig. 7 we present the maximum value of the
perturbation of each annulus as a function of the radius. The maxima
appear always along the bar (y-axis). We observe that the absolute
maximum perturbation is close to 80\% and is located a little inside
the half mass radius of the system, while the value drops to
$\approx 25\%$ near the end of the bar. For the region outside the
bar we can only trust the mean value of the saw-tooth part. This
perturbation is considered large and is often observed in real
barred galaxies. See for example the paper of \citet{b35}, where the
m=2 and m=4 components are calculated for a sample of 112 spiral
galaxies.

Observers recognise the different components of a barred galaxy by
plotting the density profile along and parallel to the the major or
minor axis of the bar as it is proposed by \citet{b32}. The bulge
length is marked by the increased light distribution over the
exponential disk well above the bar. In Fig. 8 we plot the surface
density parallel to the minor axis for two different z. For z=0
(black line) we see that a central spherical component is present
(we also mark the limits of the bar length on the x-axis). For z=1.6
$hmr$ a plateau appears that corresponds to the bar. Therefore we
can conclude that this central outcrop corresponds to a bulge that
has been created self consistently when the system has reached an
equilibrium.

Another important quantity for the orbital study is the corotation
radius. This can be determined from the maximum value of the
effective potential all along the main axis of the bar, as well as
from the equipotential contours (Fig. 9) where the $L_1$ and $L_2$
points determine approximately the radius of corotation. Therefore
we conclude that in our N-body galaxy, corotation is placed at
approximately 2.2 $hmr$. The Langrangian points $L_4$ and $L_5$ are
placed at approximately 2.0 $hmr$. Figure 9 gives extremely smoothed
contours as a result of the program used by "Mathematica".

\begin{figure}
\includegraphics*[width=7.0cm]
{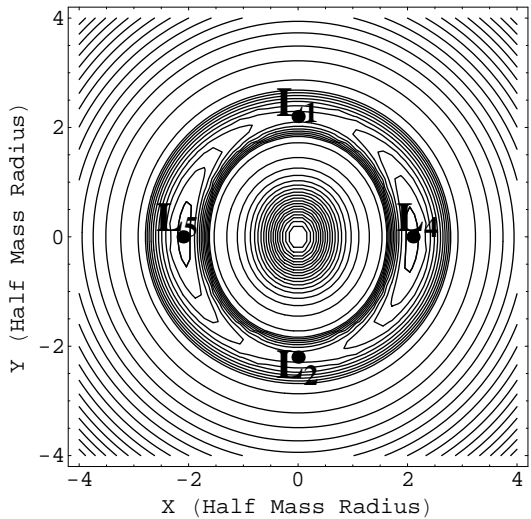}
 \caption{The equipotential contours at the end of our calculations.}
\end{figure}

\begin{figure}
\includegraphics*[width=9.0cm]
{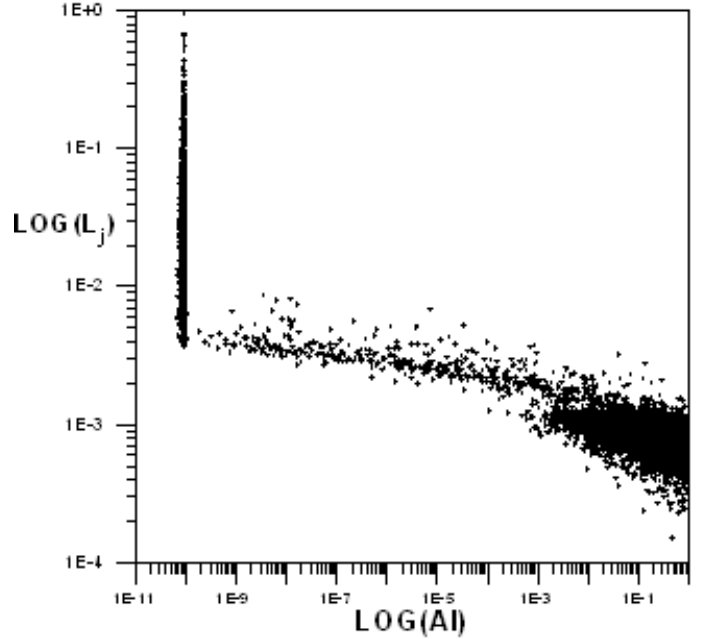}
 \caption{A snapshot of the orbits of the particles in the sample on the
plane $Log(AI)-Log(L_j)$ at 1200 radial periods.}
\end{figure}

\section{Study of the level of chaos}

The purpose of this section is to classify the orbits of our
self-consistent system in ordered and chaotic ones and find their
energy distribution.

This task can be accomplished by using a combination of two methods,
that is the Specific Finite Time Lyapunov Characteristic Number
(SFTLCN), or simply $L_j$, and the Smaller ALignment Index, (SALI),
or simply Alignment Index (AI)(Skokos 2001, Voglis et al. 2002).

The well known Finite Time Lyapunov Characteristic Number defined by
the relation

\begin{equation}
FTLCN=\frac{1}{t}log\frac{|\vec{\xi(t)}|}{|\vec{\xi(0)}|}
\end{equation}
is not very suitable for detecting chaotic orbits, because chaos is
weak in the present case and the LCN of most of these orbits is
small compared with the inverse of the dynamical time of the system;
therefore these chaotic orbits cannot be detected by integrating for
times equal to only a Hubble time.

This problem can be overcome by using the Specific Finite Time
Lyapunov Characteristic Number ($SFTLCN$), or simply $L_j$ for every
orbit which is given by the formula:

\begin{equation}
L_j(T_{rj},t_j)={T_{rj} \over  t_j} \sum_{i=1}^{N_j}a_{ij}
\end{equation}
where $t_j$ is the integration time, $N_j$ is the number of time
steps ($\Delta t=t_j/N_j$)  and $a_{ij}$ is the stretching number
(Voglis and Contopoulos 1994) at the time step $i$ ($i=1,...N_j$).
The stretching number $a_{ij}$ is defined by the equation
\begin{equation}
a_{ij}=\ln{|\xi_j(t_i + \Delta t)|  \over |\xi_j(t_i)| }
\end{equation}
where $\xi_j(t_i)$ is the length of the deviation vector from the
orbit $j$ at time $t=t_i$. To avoid numerical overflows the
deviation vectors are normalized to unity at regular time steps. The
equations giving the deviation vectors are linear and therefore a
change of their measure by a constant does not affect the
information we get.
 The time evolution of the deviation vectors
is found from the variational equations of motion. The coefficients
of these equations are evaluated numerically.
\begin{figure*}
\includegraphics*[width=17.0cm]
{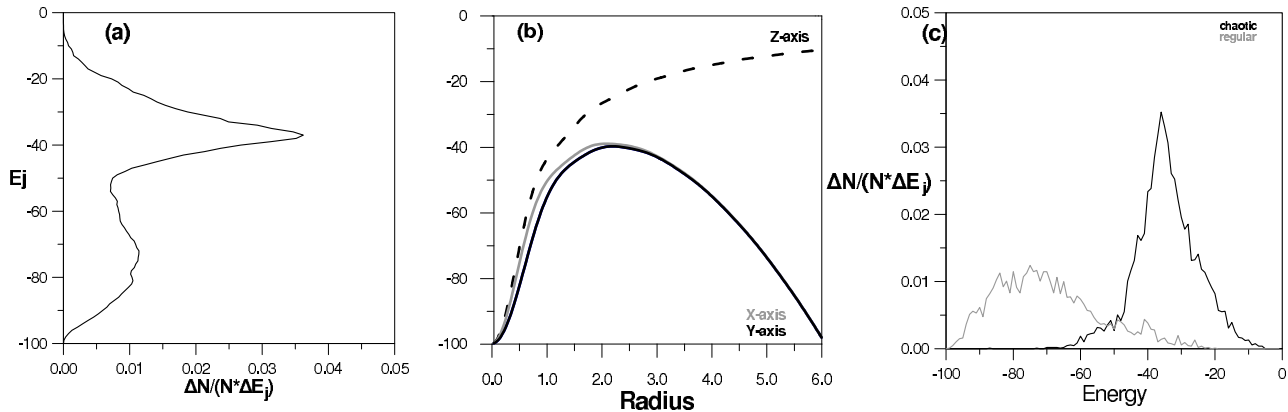}
 \caption{(a) The energy distribution of the total number of
 particles. (b) The effective potential along the X,Y and Z axes. (c) The energy distribution of
particles having chaotic orbits(black line) and of particles having
regular orbits(grey line).}
\end{figure*}
 The quantity $T_{rj}$ is the average
radial period of the orbit of the particle $j$, i.e.the time that a
star needs to go from its pericenter to its apocenter and back. This
period scales with the energy $E_j$ of the orbit as $T_{rj}
\approx\mid E_j\mid ^{-3/2}$ almost independently of the angular
momentum (Voglis 1994).

 The above definition of the SFTLCN allows us
to extend our calculations to different times for every orbit, up to
a point when the integration time is long enough to allow LCNs to be
stabilized at constant small values.

In the second method (using AI), we use the properties of the time
evolution of the deviation vectors (Voglis et al. 1998, 1999).

In particular, we consider the time evolution of two arbitrary
different initial deviation vectors $\xi_{j1}$ and $\xi_{j2}$ of the
same orbit. If the orbit is chaotic, then the smallest of the two
norms $d_{j-}(t)=|\vec{\xi}_{j1} (t)-\vec{\xi}_{j2} (t)|$ and
$d_{j+}(t)=|\vec{\xi}_{j1} (t)+\vec{\xi}_{j2} (t)|$ is called
Alignment Index (AI) and tends exponentially to zero. However, in
our calculations we impose a cutoff limit e.g. $AI=10^{-10}$ to save
integration time. On the other hand, if the orbit is ordered, then
$d_{j-}$ or $d_{j+}$ oscillate around a roughly constant mean value.
This value is usually close to unity and in any case it is not less
than $10^{-3}$.

Thus, by following the evolution of the smaller of the two indices
$d_{j-}$, $d_{j+}$, we can distinguish between regular and chaotic
orbits.

We adopt a value for the Hubble time, equal to 100 half mass
crossing times $T_{hmct}$ of the system. We usually calculate the
$L_j$s and AI of ordered orbits for the same number of radial
periods, namely $t_j/T_{rj}=1200$. However, in the case of the
chaotic orbits the integration time varies due to the cutoff limit
imposed.

The comparison of the two methods mentioned above can be shown in
Fig. 10. The sharp line on the left part of the figure (at
AI=$10^{-10}$) represents the most chaotic orbits and it is due to
the cutoff limit. The triangular distribution on the lower right
part of the figure represents the ordered orbits. Finally, the lane
of points between these two regions represents orbits that are
weakly chaotic.

 Almost 60\% of the orbits were found chaotic. This
percentage is higher than the percentage of chaotic orbits in non
rotating systems, representing elliptical galaxies, which is near
30\% at most (see for example Voglis et al. 2002). Furthermore, the
Lyapunov characteristic numbers of the chaotic orbits are larger by
about one order of magnitude than in the non-rotating
self-consistent systems, a result that is related to the resonance
effects near corotation. However only the 35\% of the total mass can
eventually develop chaotic diffusion in a Hubble time (Voglis et al.
2006).

In Fig. 11 we plot the energy distribution of the particles at the
end of our calculations, that is after 250 half mass crossing times
$T_{hmct}$ of the system. Figure 11a shows the distribution of the
total number of particles, where $\Delta E_j=1$ (in our units which
are normalized to -100 at the potential well) and represents the
amplitude of the energy bin. The distribution presents two peaks.
Figure 11b shows the effective potential along the X,Y and Z axes in
half mass radius units ($rhm$) and Fig. 11c shows the energy
distribution of chaotic orbits alone (black line) and ordered orbits
(gray line). By comparing these figures we conclude that the
sharpest peak of the distribution of the particles corresponds
mostly to chaotic orbits (since there are almost no regular orbits
there). This peak is located at $E_j\approx-36$, around corotation,
as expected, and corresponds to a radius $\approx2.2 rhm$. The
second peak is located at $E_j\approx-75$, is related exclusively to
regular orbits.

\section{Study of the orbital structure}

The purpose of this section is to classify the orbits of our
self-consistent system in such a way as to determine the shape and
the percentage of the orbits that support the bar. For this purpose
we have used two powerful tools: a) the surface of section for test
particles as well as for real N-body particles and b) the frequency
analysis of the real N-body orbits.

\subsection{Surface of
section}

We fix the values of the coefficients of the analytical expansion of
the potential and construct surfaces of section (SOS) for different
energies $E_j$ in the rotating frame (Jacobi constants):

\begin{equation}
E_j=\frac{1}{2}(v_x^2+v_y^2+v_z^2)+V(x,y,z)-\frac{1}{2} \Omega_p^2
R_{xy}^2
\end{equation}
where V(x,y,z) is the frozen potential of both the axisymmetric part
and the bar perturbation, given by the N-body run, $v_x, v_y, v_z$
are the velocities in the rotating frame of reference and $\Omega_p$
is the angular velocity of the bar rotating as a solid body and
calculated as in Fig. 3.

\begin{figure*}
\includegraphics*[width=13.cm]
{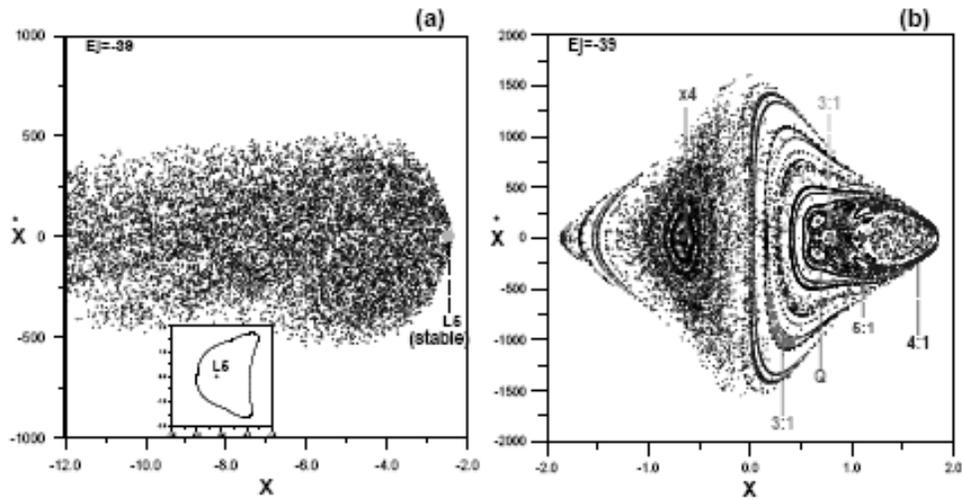} \caption{(a) The surface of section of test particles
with a value of the Jacobi constant equal to -39 for the region
outside corotation. The insert box gives the stable periodic orbit
around L5 point. (b) The same for the region inside corotation.}
\end{figure*}

We first use test particles for constructing SOS. In our 3-D model
the SOS are 4D and we plot their projections on the $x,\dot{x}$
plane, for $y=0$,  $\dot{y}>0$ and having initially $z=0$ and
$\dot{z}=0$.  All sections are made in the rotating frame of
reference. A similar technique has been used by \citet{b9} and
\citet{b13} in order to examine the phase space structure and the
foliation of the invariant tori of the orbits due to the third
integral in non-rotating triaxial self-consistent systems.

The values of $x$ are normalized with the half mass radius ($hmr$).
The Lagrangian points $L_1,L_2$ (and therefore the corotation
radius) are located at $\approx2.2 hmr$.

In Fig. 12 we plot the SOS which corresponds to a Jacobi constant
$Ej=-39$. All values of the Jacobi constant are normalized so that
the well of the potential corresponds to -100. All the initial
conditions in Fig. 12 have been chosen along $\dot{x}$=0 and they
have been integrated for 200 iterations each. Figure 12a presents
the region outside corotation where a few stable orbits around $L_5$
are present. Most of the orbits of this figure reach large distances
and a part of them escape. Figure 12b presents the region inside
corotation. Some important resonances are marked, i.e.the 3:1, 4:1
and 5:1 resonances (that appear near the end of the bar), as well as
the "x4" type orbit (Contopoulos and Papayanopoulos 1980) which has
a main axis perpendicular to the bar and is retrograde with regard
to the sign of $\Omega_p$. In Fig. 13 the corresponding periodic
orbits are plotted. Every periodic orbit intersects the SOS at one
or a finite number of points. The stable periodic orbits are easily
identified on the SOS, because they are surrounded by closed
invariant curves.

\begin{figure*}
\includegraphics*[width=11.cm]
{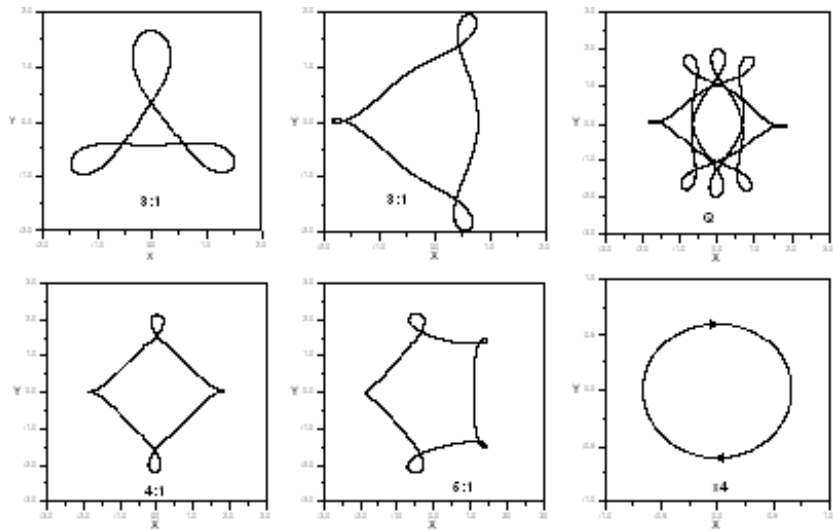} \caption{Periodic orbits that intersect the SOS at the
points indicated in Fig. 12(b).}
\end{figure*}
\begin{figure*}
\includegraphics*[width=15.0cm]
{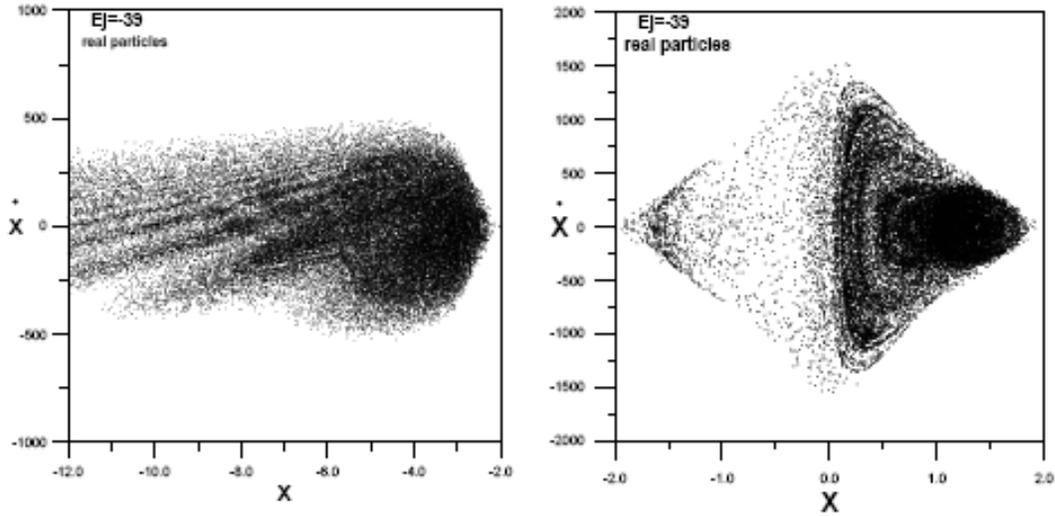} \caption{The surface of section of real N-body particles
with a value of the Jacobi constant equal to -39: (a) for the region
outside corotation and (b) for the region inside corotation.}
\end{figure*}

\begin{figure*}
\includegraphics*[width=14.0cm]
{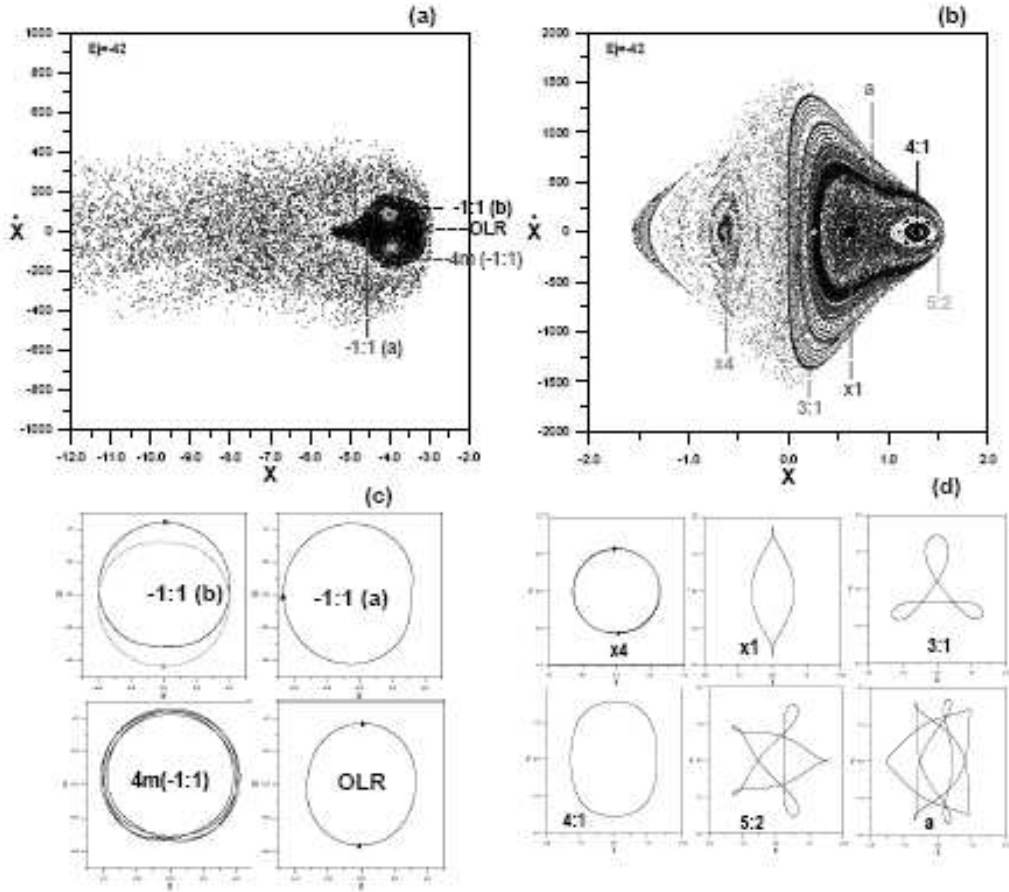} \caption{The surface of section of test particles with a
value of the Jacobi constant equal to -42: (a) for the region
outside corotation and (b) for the region inside corotation. (c)
Orbits corresponding to the region outside corotation: There are two
orbits (-1:1 (b)) and a multiplicity 4 orbit that has bifurcated
from -1:1(b). (d) Orbits corresponding to the region inside
corotation.}
\end{figure*}
\begin{figure*}
\includegraphics*[width=17.0cm]
{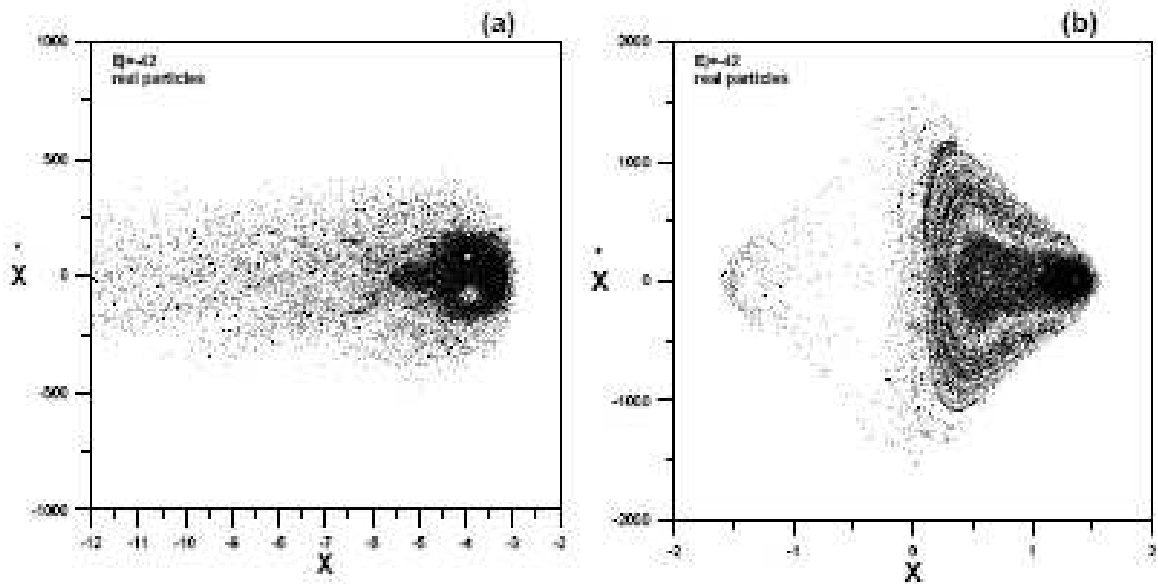} \caption{Same as in Figure 15a,b but for the real N-body
particles: (a) for the region outside corotation and (b) for the
region inside corotation.}
\end{figure*}
\begin{figure*}
\includegraphics*[width=12.cm]
{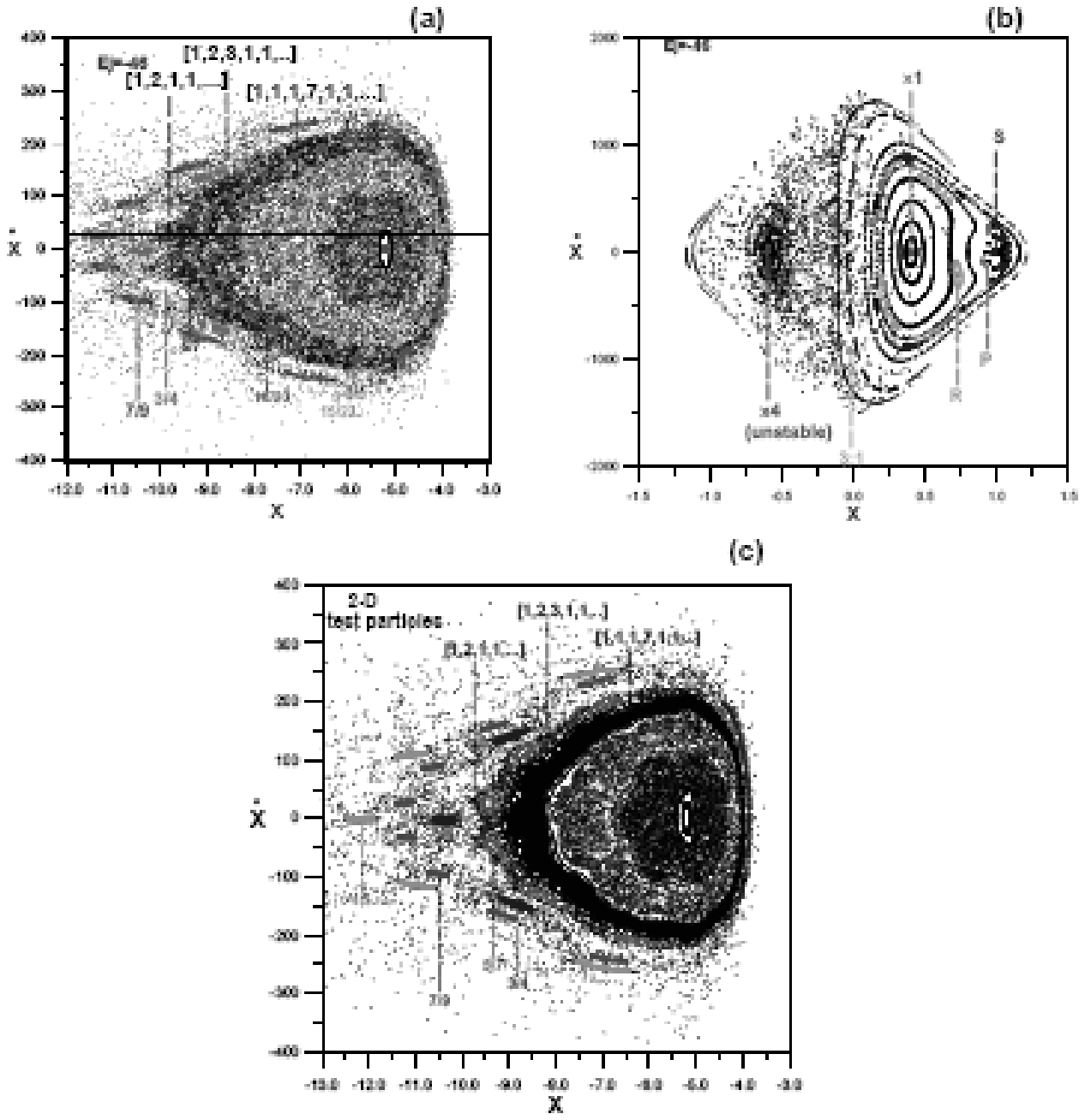} \caption{SOS for $E_j=-46$: (a) for test particles for
the region outside corotation and (b) for the region inside
corotation. (c) SOS for the 2-D approximation of the region outside
corotation.}
\end{figure*}

 In Fig. 14 we
have plotted the real particles from the N-body run. We have chosen
the initial conditions of particles having a Jacobi constant equal
to $E_j=-39\pm0.5$ and we have integrated them in a fixed potential
for 100 iterations. In this energy level only a small fraction of
the orbits of real particles are regular as it can be seen from Fig.
11c. Figures 12a and 14a present the region outside corotation
containing test particles and real N-body orbits respectively around
the $L_5$ point. We see several thick dark lines that are called
"rays". An explanation of the ray structure of this figure was given
by \citet{b10} who noticed that even in the chaotic region we may
see a structure of rays (see Fig. 9 of \citet{b10}). The successive
iterates of every initial point along a ray are located also on
rays, and they move in a clockwise way around the center of the
black region. After each return the points deviate from their
initial positions and therefore the rays thicken in time and finally
they join each other. Thus the ray structure disappears if we take a
much greater number of iterations than in Fig. 14a.

 Figure 14b presents
the region inside corotation. It is obvious that the area
corresponding to the "x4" orbits is depopulated while the orbits
 supporting the shape of the bar and having the same sense of rotation with it,
are well populated.

In Fig. 15 the surface of section for test particles is plotted for
another value of the Jacobi constant $E_j=-42$. Figure 15a presents
the region outside corotation. It is remarkable that some islands of
stability exist that correspond to the OLR and -1:1 resonances.
Their corresponding orbits are shown in Fig. 15c. For the area
inside corotation (Fig. 15b) the stable orbits correspond to the
3:1, 4:1 and 5:2 resonances, as well as to the "x1" type of orbits
with loops at the edges and the retrograde "x4". In Fig. 15d the
corresponding orbits are plotted.

The same surface of section for the real N-body particles is shown
in Fig. 16. The main resonances supporting the bar are populated
here again while the area around the "x4" orbit is depopulated.
Another important remark is that real particles avoid the area of
the OLR as well as the -1:1 resonance outside corotation. A
remarkable feature common for test particles as well as for real
N-body particles is the intensely dark area around the OLR and -1:1
resonances in the region outside corotation. (compare Figs. 14a and
16a). This is an indication of the existence of a cantorus which
impedes the communication of the two chaotic regions, imposing a
partial barrier in this 2-D projection of the section. On the other
hand since our N-body system is 3-dimensional, one would expect that
Arnold diffusion would transport particles through the third
dimension and produce a communication between the two chaotic
regions (for a definition of Arnold diffusion see Contopoulos 2002).
However, our results indicate that such a communication is very
slow, therefore Arnold diffusion is not efficient.
\begin{figure*}
\centering
\includegraphics*[width=14.0cm]
{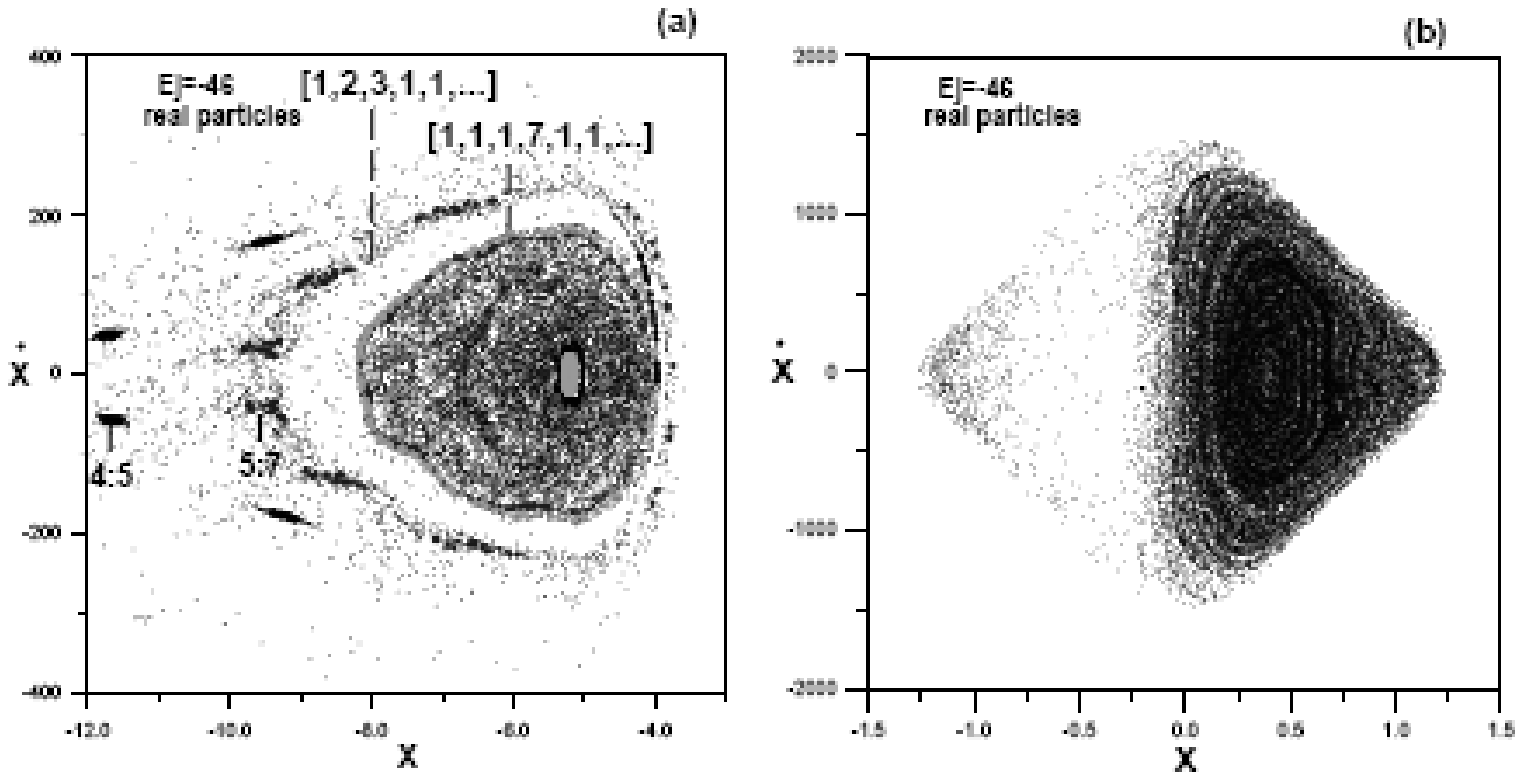} \caption{Same as in Fig. 17a,b for the real N-body
particles.}
\end{figure*}
In Fig. 17 the SOS is plotted for test particles, with $E_J=-46$ for
the region outside corotation (Fig. 17a) and for the region inside
corotation (Fig. 17b). Figure 17c shows the 2-D approximation of the
SOS when there is no third dimension, i.e. the forces on the z-axis
are exactly zero. In this figure we mark some important tori. Every
torus is characterized by its rotation number,i.e. the average angle
between successive intersections of an orbit by the SOS, as seen
from a central fixed point. This number can be represented in the
form of a continued fraction:
\begin{equation}
[\alpha_0,\alpha_1,...]\equiv\frac{1}{\alpha_0+\frac{1}{\alpha_1+...}}
\end{equation}
If $\alpha_i=1$ for all i beyond a certain j, this number is called
a noble number.

In the region outside corotation the existence of different tori and
cantori separates the different chaotic regions. By comparing Fig.
17c with Fig. 17a we see only minor differences. In Fig. 17c the
chaotic region
 inside the torus [1,1,1,7,1,1,..] is united but it is
obvious that a cantorus still exists separating the two unequally
darkened areas (a cantorus is a torus with a Cantor set of holes).
We have taken an initial condition in the most darkened area of this
SOS, between [1,1,1,7,1,1,..] and [1,2,3,1,1,..] and we have seen
that the orbit stays located there for at least 5x$10^{5}$
iterations and possibly for ever. Therefore we conclude that there
may still exist tori (or cantori with very small holes)
corresponding to a noble number [1,1,1,7,1,1..] that puts an inside
barrier and to [1,2,3,1,1,..] that puts an outside barrier for this
chaotic area.

The same phenomenon is obvious in Fig. 18 where the same projection
of SOS is plotted for real N-body particles for the region outside
corotation (a) and for the region inside corotation (b). Here again
it is obvious that for the region outside corotation, the real
orbits (that are chaotic in their majority) are restricted in areas
that are bounded by the same tori, while there are areas with almost
no N-body particles at all. Therefore tori and cantori on the
projection of the SOS, play an important role in retaining chaotic
orbits projected on the 2-D space of the galaxy, limiting the
effectiveness of Arnold diffusion through the third dimension.

The range of energies for which there exist cantori as described
above is -60$<E_j<$-45. In this range the amounts of ordered and
chaotic orbits are of the same order (see Fig. 11c).

There are two mechanisms that force chaotic orbits to stay bound
inside specific areas and make them dynamically important for the
system. The first one, that has been described above, is the
phenomenon of cantori that put a partial barrier on the SOS and
reduce the effectiveness of Arnold diffusion. The second one is the
stickiness effect of a fraction of the chaotic orbits near
resonances that support the bar. Stickiness appears in systems of 2
or more degrees of freedom when chaotic orbits remain close to an
invariant manifold for long times before escaping to large
distances. This phenomenon is described in the book of \citet{b36}.
This can be shown from the frequency analysis in the next section.
The latter mechanism makes the role of the chaotic layer near the
limits of the bar important in supporting its ellipticity and
boxiness.

In Fig. 19 we plot the orbits corresponding to islands of stability
for the region inside corotation of Fig. 17b. The rotation number is
the average frequency of rotation around a fixed point of a stable
periodic orbit (here OLR) on the phase space. The rotation number as
a function of x along the thick line shown in Fig. 17a is plotted in
Fig. 20. In this figure we mark several islands of stability with
rational rotation numbers and the noble numbers of some important
cantori. We note that the number of islands is smaller than in the
case of \citet{b10}. This is due to the fact that the bar
perturbation is larger for the corresponding Jacobi constant in the
present model, thus chaos is stronger.

\subsection{Frequency analysis of the orbits}

In Fig. 21 we plot the main frequencies of all the regular orbits
out of a sample of 10764 particles at the end of the N-body run in
polar coordinates. We have used time series for each orbit that
correspond to a time interval of 100 radial periods. The abscissa
corresponds to the frequency of the variation of the radius of the
orbit on the plane of rotation (x-y plane in the rotating frame of
reference)
 and the ordinate corresponds to the frequency of the angle $\varphi$ between the
position vector on the plane of rotation and the main axis of the
bar (y axis in the rotating frame of reference). Both frequencies
are regularized by the frequency in the third dimension (z-axis). It
is obvious from this figure that there are groups of orbits
concentrated all along lines of specific resonances and others that
are more scattered in between the lines of resonances. Moreover
along the same line there can be different groups of orbits
belonging to the same resonance. For example on the 2:1 line there
are three groups of orbits: the "x1" type with the main axis along
the axis of the bar, the "x4" type (only a very small fraction of
the total sample) with the main axis perpendicular to the main axis
of the bar, and the 2:1 orbits near corotation (n.c) that have a
more rectangular shape.

In Fig. 22 we plot the three projections of some real N-body orbits
that correspond to different resonances. From this figure it is
obvious that in the majority of the orbits that are near resonances
the third dimension is not of great importance and the projections
on the plane of rotation have shapes that support the bar. We denote
as "box" the orbits that are found scattered in between the
resonances. In particular we see some groups of orbits that are
concentrated in rather small areas in Fig. 21, like Group A and
Group B. Group A has orbits that support the bar. Orbits of this
type near the limit of the bar are boxy in shape. On the other hand
Group B orbits are more spherical on the plane of rotation and their
projections on the other planes have a considerable thickness.

Figure 23 is the same as Fig. 21 but for the chaotic orbits of our
self-consistent system. The orbits are calculated here up to 300
radial periods. The distribution is more scattered on this figure,
as expected. However we can still identify concentrations around
resonances. This is a consequence of the fact that most of these
orbits are only weakly chaotic.

\begin{figure}
\includegraphics*[width=8.cm]
{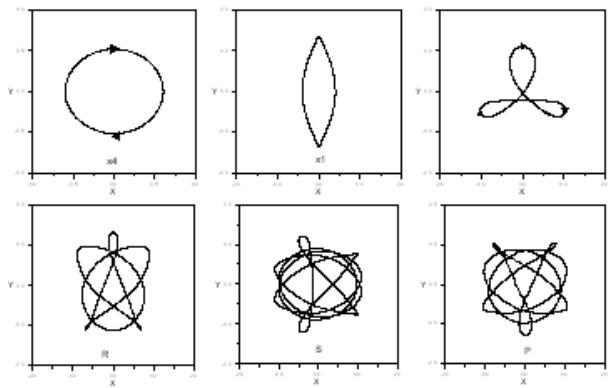} \caption{Orbits corresponding to the SOS of Fig. 17b.}
\end{figure}

In Fig. 24 we plot the projections of some real N-body chaotic
orbits that have frequencies near resonances. On the xy-plane (of
the rotating frame of reference) we have plotted the orbits for 100
radial periods as well as for 300 radial periods. We see that orbits
do not change considerably from 100 to 300 periods.

Group A corresponds to orbits with a boxy shape and can be found
near the end of the bar as well as in smaller radii. Group B
corresponds to orbits more spherical and in general with their main
axis perpendicular to the bar with an important thickness in the 3rd
direction (z-axis). Most of these orbits present a "X-shape" on the
zx plane.
\begin{figure}
\includegraphics*[width=7.cm,angle=-90]
{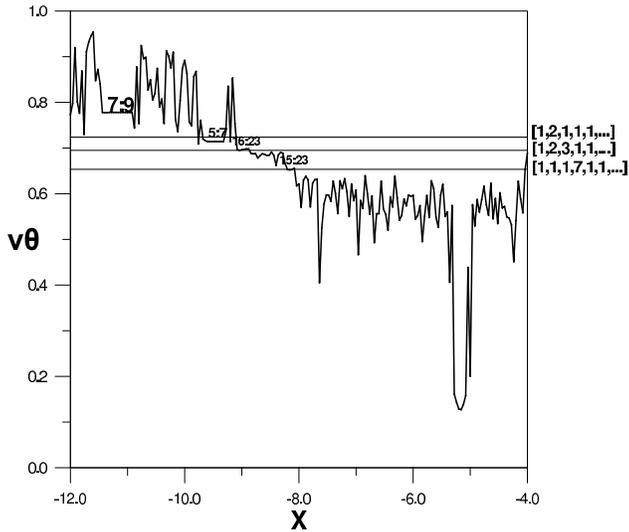} \caption{The rotation number as a function of x along
the thick line shown in Fig. 17a.}
\end{figure}

\begin{figure}
\includegraphics*[width=10.cm]
{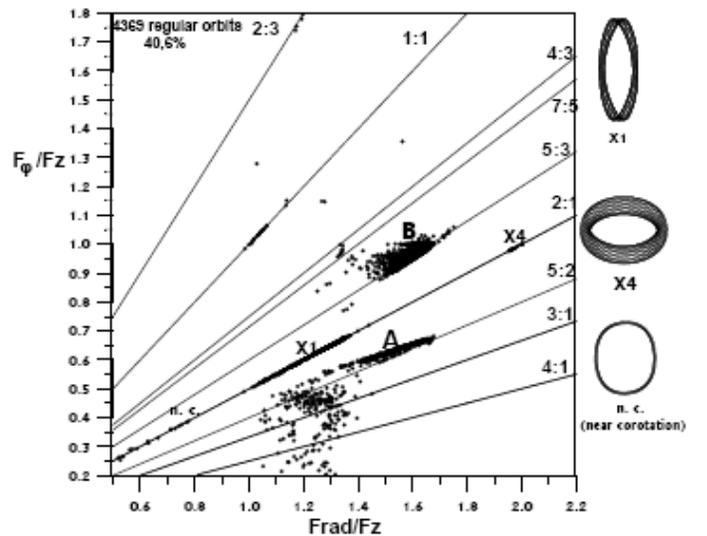} \caption{Frequency analysis for the regular orbits.}
\end{figure}

\begin{figure}
\includegraphics*[width=8.cm]
{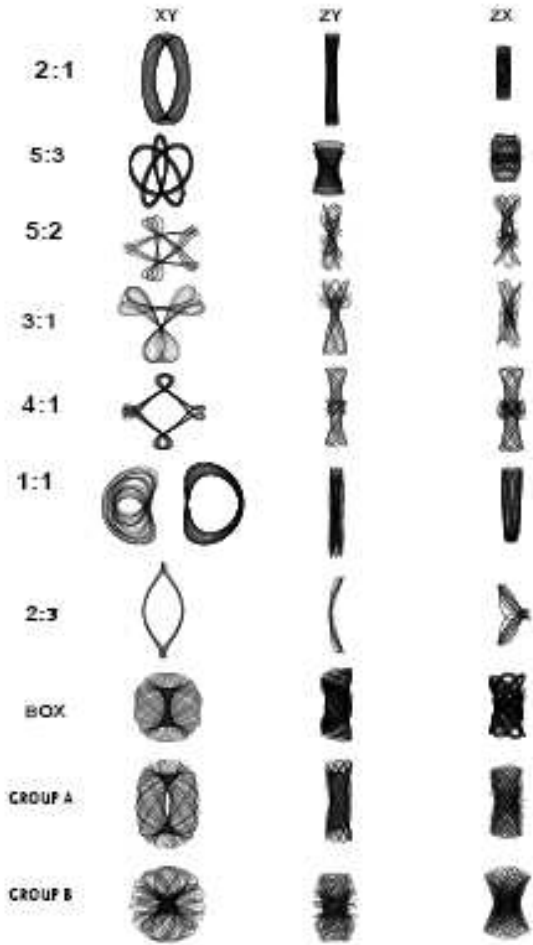} \caption{Some examples of real N-body regular orbits.}
\end{figure}
\begin{figure}
\includegraphics*[width=9.0cm]
{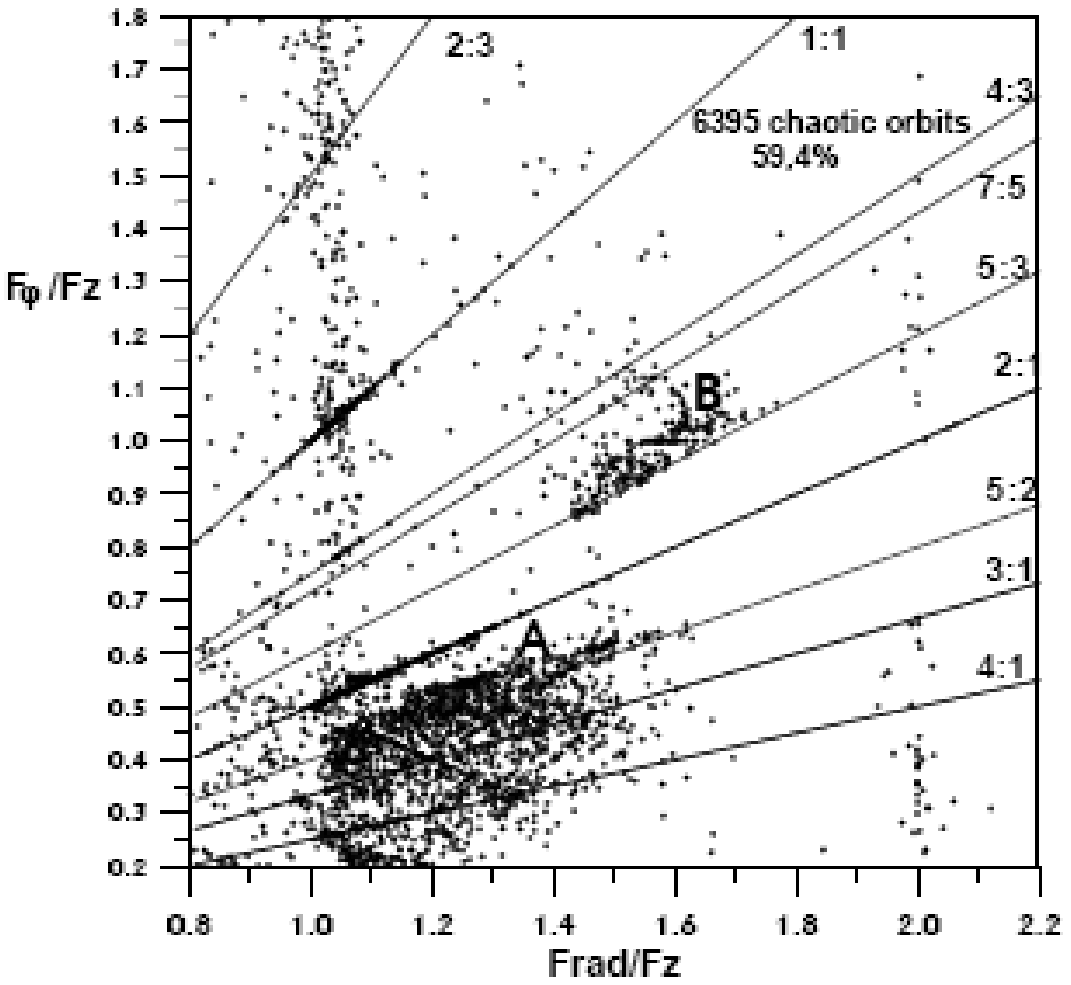} \caption{Frequency analysis for the chaotic orbits.}
\end{figure}
We have also found some families of 3-D orbits whose projections on
the rotating plane support the bar and have been found and named in
3-D models by \citet{b22}. In Fig. 24 there are examples of real
N-body chaotic orbits with boxy shape that support the bar all along
its radius, like for example the 2:1 and 4:1 resonant orbits or the
orbits of group A. Similar orbits are found in models of barred
galaxies by \citet{b16}. The spatial distribution of the chaotic
orbits is in general more spherical than the one of the regular
orbits, i.e. the shape of the bar is supported mostly by the regular
orbits (Voglis et al. 2006). However there is a layer of weakly
chaotic orbits that belong mostly to the outer part of the bar and
support the shape of the bar.

\begin{figure}
\includegraphics*[width=9.cm]
{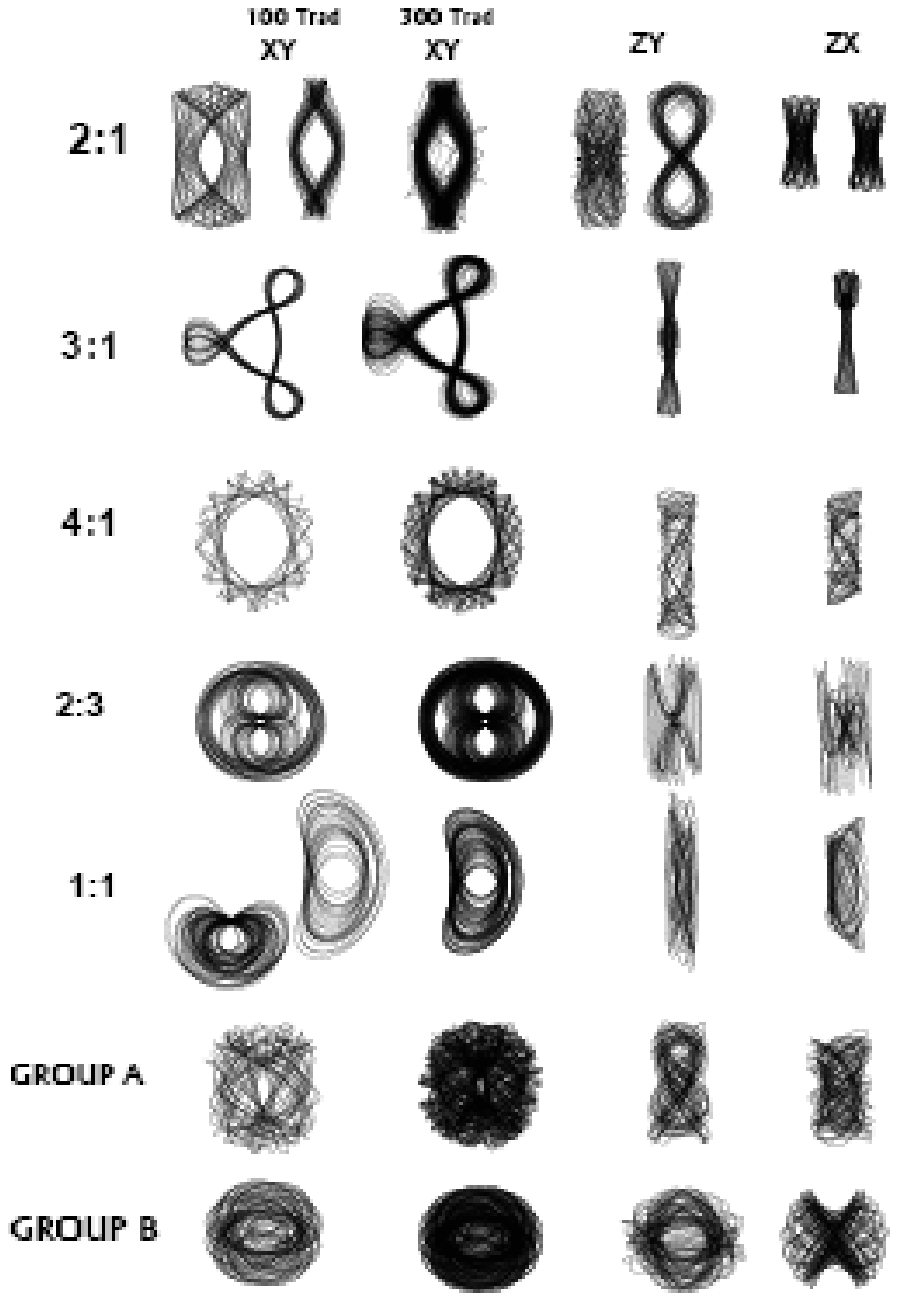} \caption{Some examples of real N-body chaotic orbits.}
\end{figure}

\begin{figure}
\includegraphics*[width=6.0cm,angle=-90]
{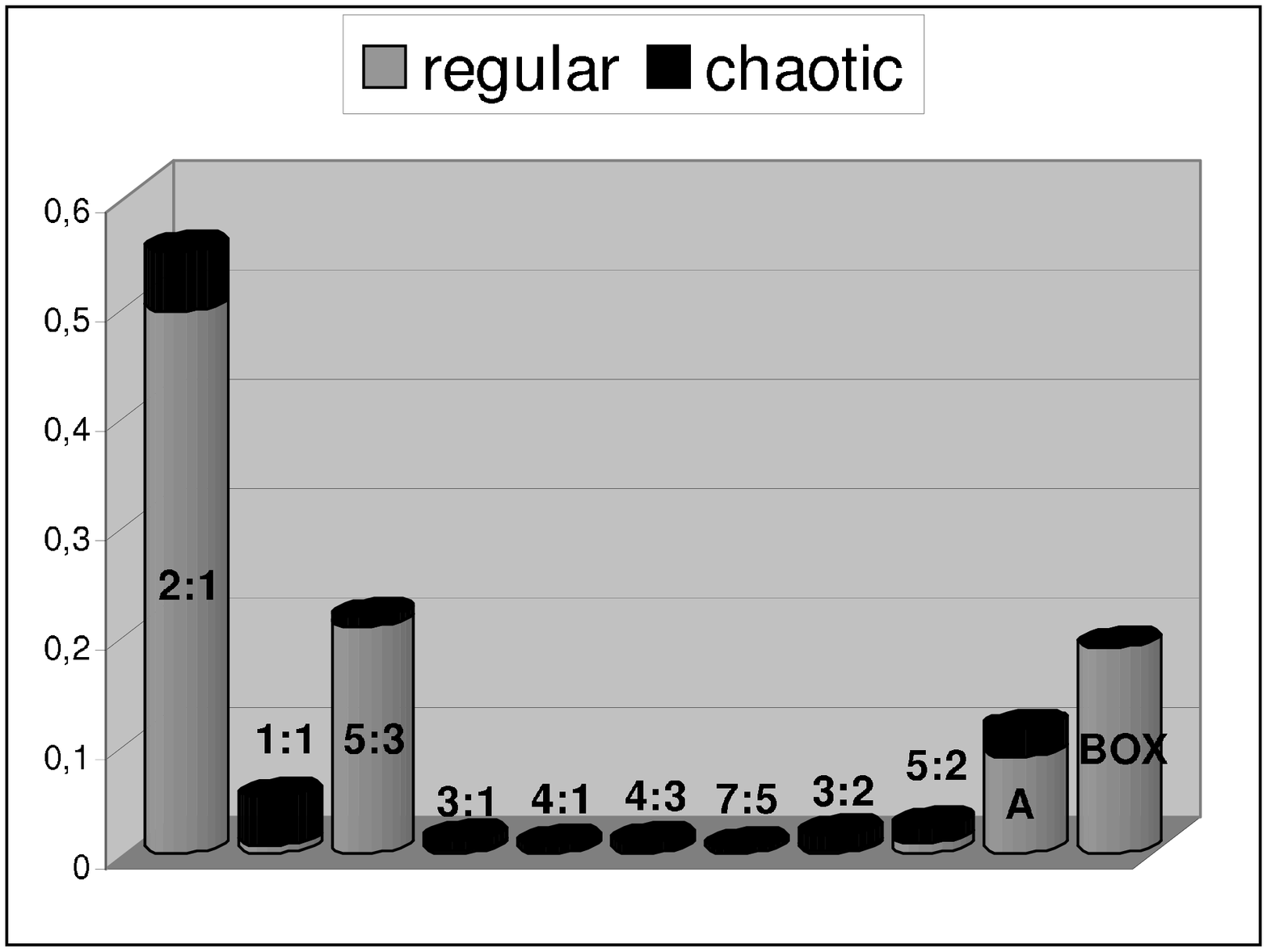} \caption{The statistics of the regular and chaotic
orbits.}
\end{figure}
In Fig. 25 we show the statistics of the resonances for regular
(gray) as well as for chaotic orbits (black). We take a width
$\pm0.02$ around each resonance to populate the orbits. We find that
20\% of the chaotic orbits lie near resonances and only orbits of
this percentage are plotted in Fig. 24.

In \citet{b31} the authors have found for this experiment (named
QR1) that although 60\% of the total matter is chaotic, only a
fraction of 35\% of the total matter can develop chaotic diffusion
in a Hubble time. Therefore the rest 25\% of the total matter is
weakly chaotic and is mostly located in a layer at the edge of the
bar in resonant form, or in regions outside corotation limited by
certain tori and cantori (see Figs. 16,17).

\section{CONCLUSIONS}

We have studied the orbital structure of rotating self-consistent
bar-like N-body equilibrium systems. The amplitude of the bar is
relatively large ($\delta \varrho/\varrho_0$ is about 80\%,
maximum). The angular velocity of rotation of the pattern is
$\Omega_p$=($2\pi$)/($17hmct$)=0.37$rad/hmct$, which is equivalent
to 8.25$km/(kpc.sec)$, the spin parameter being roughly that of our
Galaxy.

The main conclusions of our study are the following:

1) In our N-body experiment that simulates a rotating barred galaxy,
chaos is very appreciable, amounting to about 60\% of the total
bound matter in contrast with nonrotating systems, which have
$\approx 30\%$ chaos. However chaos is in general weak and only
about 35\% of the mass develops chaotic diffusion in a Hubble time.

2) The ellipticity of the bar decreases outwards from 0.5 (E5
galaxy) in the inner parts to about 0.1 (E1 galaxy) in the outer
parts.

3) The bar is like an orthogonal parallelepiped with a boxiness
parameter that is maximum in the middle of the bar.

4) Only weak spirals appear in the present model, and they vanish
relatively fast.

5) A comparison is made between systems of test particles and N-body
systems by projecting the orbits on a 2-D surface of section (SOS).
We notice, first, that  in both cases chaos is dominant outside
corotation, while most orbits in the bar are ordered.

6) In the outer regions of the SOS of test particles as well as of
real N-body particles, we see some thick dark "rays". The particles
of these rays have their images along other rays, but after longer
times these images fill the regions between the rays, and the rays
disappear.

7) Certain outer regions of the SOS have different degrees of
darkness in the case of test particles. These regions are separated
by invariant tori or cantori. The invariant tori do not allow any
communication across them in the 2-D case. In the 3-D case we see
similar results although communication is in principle possible by
means of Alnold diffusion. Thus Alnold diffusion is probably not
effective in the present case. On the other hand cantori allow
communication across them but this process is slow if the cantori
have small holes, and for a long time the areas from both sides of
these cantori do not communicate.

8)The regions inside corotation on the SOS contain mostly ordered
orbits. However there are important differences between the test
particles case and the N-body case. In particular real orbits do not
follow retrograde orbits around x4. In general N-body orbits do not
enter the regions of islands if they were not inside such islands
originally.

9) We have found the forms of the periodic (planar) orbits and the
rotation numbers along a particular straight line on the surface of
section that allows to locate the small islands and the positions of
tori and cantori. These tori and cantori have noble rotation
numbers, i.e. their expansions in continued fractions end with an
infinity of 1s.

10) A frequency analysis of ordered orbits has identified clearly
the various resonant forms of nonperiodic orbits that form islands
of stability. We found also stable orbits of box type outside the
islands.

11) We did also a frequency analysis of chaotic orbits. The
frequency diagram has several similarities with the corresponding
diagram for regular orbits, although it is more diffuse. This is due
to the fact that chaos is rather weak in the present case.

12) Finally we give the statistics of the various types of ordered
and chaotic orbits. The most important orbits are the x1 orbits, of
2:1 type. These orbits support the bar. But other orbits, like the
box orbits and the orbits of group A also support the bar. In
particular the group A orbits are mainly responsible for the
boxiness of the bar. A 20\% of the chaotic population are located
near resonances and play an important role in supporting the
boxiness of the bar in its outer layers. On the other hand there is
also a percentage of chaotic orbits that are located in restricted
areas outside corotation limited by tori and cantori on the 2-D
projections of SOS. The effectiveness of Arnold diffusion around
these tori and cantori is restricted.

In a future paper experiments with stronger and longer lived spiral
arms will be presented.

\section*{Acknowledgments}
We would like to thank Dr. P. Patsis and Dr. C. Kalapotharakos for
fruitful discussions.

\bsp

\label{lastpage}

\end{document}